\documentclass[12pt,aps,amsmath,latexsym,amsfonts,letterpaper]{JHEP3}

\usepackage{graphicx}
\usepackage{epsfig}
\usepackage[all]{xy}
\usepackage{subfigure}

\input epsf.tex

\def\half{\textstyle{1\over2}}

\def\nn{\nonumber}

\newcommand{\be}{\begin{equation}}
\newcommand{\ee}{\end{equation}}
\newcommand{\bea}{\begin{eqnarray}}
\newcommand{\eea}{\end{eqnarray}}
\newcommand{\bml}{\begin{mathletters}}
\newcommand{\eml}{\end{mathletters}}

\renewcommand{\d}{\ensuremath{{\rm d}}}

\newcommand{\ba}{\begin{eqnarray}}
\newcommand{\ea}{\end{eqnarray}}

\newcommand{\lab}{\label}

\newcommand{\beq}{\begin{equation}}
\newcommand{\eeq}{\end{equation}}
\newcommand{\beqa}{\begin{eqnarray}}
\newcommand{\eeqa}{\end{eqnarray}}
\newcommand{\beqar}{\begin{eqnarray*}}
\newcommand{\eeqar}{\end{eqnarray*}}

\title{The evolution of conifolds.}
\author{Neil A. Butcher\footnote{email: ppxnb@nottingham.ac.uk}  and 
        Paul M. Saffin\footnote{email: paul.saffin@nottingham.ac.uk}\\
School of Physics and Astronomy, University Park, University of
Nottingham, Nottingham NG7 2RD, UK}

\date{\today}

\maketitle

\abstract{
We simulate the gravitational dynamics of the conifold geometries (resolved and deformed) involved in the description of certain compact spacetimes. As the cycles of the conifold collapse towards a singular geometry we find that a horizon develops, shielding the external spacetime from the curvature singularity of the newly formed black hole. The structure of the black hole is examined for a range of initial conditions, and we find a candidate black-hole solution for the final state of the collapse.
}

\preprint{arXiv:yymm.nnnn [hep-th]}

\begin{document}

%%%%%%%%%%%%%%%%%%%%%%%%%%%%%%%%%%%%%%%%%%%%%%%%%%%%%
\section{Introduction}
\lab{intro}
%%%%%%%%%%%%%%%%%%%%%%%%%%%%%%%%%%%%%%%%%%%%%%%%%%%%%
String theories present us with a natural opportunity to study extra dimensions. Within the Kaluza-Klein picture of the Universe these extra dimensions take the form of small, compact spaces
with the full spacetime comprising four large dimensions with which we are acquainted and a compact manifold, ${\cal M}$. While the large dimensions can be observed, the nature of the small manifold ${\cal M}$ cannot be directly seen, however some properties of ${\cal M}$ can be deduced such as its
dimension and, with certain assumptions, its topology.
Provided there are no D-branes or fluxes, a situation we shall not consider in this paper, the manifold ${\cal M}$ must have special holonomy in order to preserve supersymmetry in the low energy theory\cite{Candelas:1984yd}. This restriction on the holonomy group implies that the manifold is a Calabi-Yau manifold. However there are a great many six dimensional Calabi-Yau manifolds and any one could be used for the compactification. Within each topology class there can be a further choice for the geometry, given by the value of the moduli parameters, where spaces with the same topology are grouped into moduli spaces and the continuous degeneracy of the moduli leads to massless fields in the low energy theory, moduli fields. Selecting one manifold to compactify on, given the huge range of possibilities, is a significant challenge for string theory; the observational consequences depend very heavily on ${\cal M}$, and without a unique manifold we lack the unique predictions we require.

This situation was improved when it was realized that many of these Calabi-Yau manifolds were connected in moduli space\cite{Candelas:1988di,Candelas:1989ug,Greene:1996cy}, with the connecting geometry being singular. One way to picture these transitions between manifolds is to study the cycles within them, for example it may be that certain cycles collapse to zero size on one side of the transition and expand as different cycles on the other. Transitions may change the Hodge numbers, such as a conifold transition \cite{Candelas:1989js,Gwyn:2007qf} where an $S^2$ collapses and reappears as an $S^3$ or they may maintain the Hodge numbers but alter the linking numbers of the cycles, such as the flop transition \cite{Greene:1996cy}. Remarkably, string theory is able to make sense of these singular geometries by the appearance of new light states corresponding to D-branes wrapping the collapsing cycles \cite{Strominger:1995cz}. We can also investigate the low energy theory\cite{Greene:1996dh} and deduce its dynamics\cite{Lukas:2004du,Palti:2005kv,Mohaupt:2004pq,Mohaupt:2004pr,Mohaupt:2005pa}.
The singular geometry has acted to connect the moduli spaces, its singularity is conical and takes the local description of a discrete quotient of a smooth manifold $\tilde{X}$, $X=\tilde{X}/\Gamma$, where $\Gamma$ is a finite symmetry group; the singularity is then the fixed point set of $\Gamma$. Of particular interest are those singularities which allow a smooth resolution, by blowing up certain cycles of zero size contained within the fixed point set of $\Gamma$. This is achieved in practice by a surgery which replaces a ball around the conical singularity with a ball of a smooth special-holonomy space \cite{Gibbons:1979xn,Page:1979zu,Joyce:2000}. The moduli of this special-holonomy space then appear in the low-energy theory as moduli fields \cite{Lukas:2003dn}.

We investigate the dynamics of the internal space while the cycles collapse, as in \cite{Butcher:2007zk} we are interested in the gravitational effects surrounding the collapsing cycles, most notably whether a horizon forms to surround the cycle. The appearance of a horizon in the spacetime would indicate that the process acting to change the topology cannot be so easily understood as was previously thought, with the horizon acting to void the use of the currently-understood low energy theory close to the conical singularity. Previous studies have used the low energy dynamics of the moduli fields and inferred the results of the topology change from these; the formation of a horizon invalidates this approach.
We will consider two geometries which act to change their topology by means of a conifold transition, namely the resolved and deformed conifods. We introduce these two spaces in the next section along with the nature of the transition we wish to instigate. We go on in sections \ref{sec:ResCol} and \ref{sec:DefCol} to describe the simulations we performed along with their results. We then attempt to ascertain the nature of the higher dimensional black hole in section \ref{sec:blackhole}, and go on to give our conclusions in section \ref{sec:Conc}.

%%%%%%%%%%%%%%%%%%%%%%%%%%%%%%%%%%%%%%%%%%%%%%%%%%%%%
\section{Conifolds.}
\lab{sec:Conifold}
%%%%%%%%%%%%%%%%%%%%%%%%%%%%%%%%%%%%%%%%%%%%%%%%%%%%%

The conifold is a gravitational instanton exhibiting a singular point, the position of which we choose to be the origin. The utility of this space is that it represents the generic singularity within a class of Calabi-Yau gemetries \cite{Greene:1995hu}, and we know the explicit metric for the singular geometry \cite{Candelas:1989js} and its two smooth versions. These other two manifolds, called the deformed conifold and the resolved conifold, are similar to the singular conifold in that they have the same asymptotic structure, and they all admit a k\"ahler metric. However, they differ from the conifold in that the singular point has been replaced in some way. The resolved conifold has an $S^2$ as a replacement for the tip, whereas the deformed conifold exhibits an $S^3$. Following are the three metrics written within a specified coordinate system.

%%%%%%%%%%%%%%%%%%%%%%%%%%%%%%%%%%%%%%%%%%%%%%%%%%%%%
\subsection{Singular conifold}
\lab{sec:sing}
%%%%%%%%%%%%%%%%%%%%%%%%%%%%%%%%%%%%%%%%%%%%%%%%%%%%%

A singular conifold is a Calabi-Yau threefold, it takes the form of a cone with a five-dimensional $T^{(1,1)}$ base,

\be
\lab{eq:ConMetric}
ds^2=dr^2+r^2 (ds_{base})^2.
\ee
\cite{Candelas:1989js,Gwyn:2007qf}.

Such a base has the form of $S^3$x$S^2$\cite{Candelas:1989js} and can be written in terms of two $SU(2)$ groups, using two distinct sets of the conventional left-invariant one-forms of $SU(2)$ which satisfy
\bea
\lab{eq:2forms}
\nn\d\sigma^i&=&-\half\epsilon_{ijk}\sigma^j\wedge\sigma^k,\\
\d\Sigma^i&=&-\half\epsilon_{ijk}\Sigma^j\wedge\Sigma^k.
\eea

With these we can write the base of the conifold
\bea
\lab{eq:ConBase}
d(s_{base})^2=\frac{1}{6}(\sigma_1^2+\sigma_2^2+\Sigma_1^2+\Sigma_2^2)+\frac{1}{9}(\sigma_3+\Sigma_3)^2.
\eea

%%%%%%%%%%%%%%%%%%%%%%%%%%%%%%%%%%%%%%%%%%%%%%%%%%%%%
\subsection{Resolved conifold}
\lab{sec:Resolved}
%%%%%%%%%%%%%%%%%%%%%%%%%%%%%%%%%%%%%%%%%%%%%%%%%%%%%

The resolved conifold is the alteration to the singular conifold achieved by expanding the singular point into an $S^2$\cite{Pando Zayas:2000sq,Candelas:1989js}. It is further defined by a parameter determining the radius of the $S^2$, we call this parameter $\alpha$. In the limit that this radius drops to zero, we recover the singular conifold. Using the same one-forms to represent $SU(2)$ the resolved conifold can be expressed as follows.

\bea
\lab{eq:resMetric}
ds^2\nn=-dt^2&+&\frac{r^2+6\alpha^2}{r^2+9\alpha^2}dr^2+\frac{1}{6}r^2(\sigma_1^2+\sigma_2^2)\\&+&\frac{1}{6}(6\alpha^2+r^2)(\Sigma_1^2+\Sigma_2^2)+\frac{r^2}{9}\left(\frac{r^2+9\alpha^2}{r^2+6\alpha^2}\right)(\sigma_3+\Sigma_3)^2.
\eea

At the origin this metric does not degenerate to a point, instead we find an $S^2$ which can be seen by setting r=0,
\be
ds^2|_{r=0}=-dt^2+\frac{6}{9}dr^2+\alpha^2(\Sigma_1^2+\Sigma_2^2).
\ee

This is a 2-sphere of area, $4\pi\alpha^2 $ (crossed with a 1+1 Minkowski geometry) which has replaced the origin of the singular conifold, and in doing so has smoothed the manifold.
Clearly the resolved conifold approaches the singular conifold in either the high $r$ or low $\alpha$ limits.

%%%%%%%%%%%%%%%%%%%%%%%%%%%%%%%%%%%%%%%%%%%%%%%%%%%%%
\subsubsection{Moduli space approximation}
\lab{sec:Moduli}
%%%%%%%%%%%%%%%%%%%%%%%%%%%%%%%%%%%%%%%%%%%%%%%%%%%%%
In order to estimate the low energy dynamics of the resolved conifold we can allow the modulus $\alpha$ to vary in time to a small extent \cite{Manton:1981mp}. This involves introducing a new, time dependent modulus, which we called $\alpha(t)$ where $\alpha(0)=\alpha$. If we use the Einstein-Hilbert action and the resolved conifold metric, including this new dependence, we get an effective action:
\be
\label{eq:modAction}
S_{eff}=\int dt\,\left(\frac{d}{dt}\sqrt{\alpha}\right)^2.
\ee

This gives the approximation that $\sqrt{\alpha(t)}$ is linear in time, implying that if we set $\alpha(t)$ on a course towards hitting zero, then the moduli approximation says that it will reach zero in finite time. As $\alpha=0$ corresponds to the singular geometry then we see no obstruction to the geometry becoming singular within this approximation. By extending this approximation to include the full gravitational dynamics we hope to achieve a better understanding of this process.

%%%%%%%%%%%%%%%%%%%%%%%%%%%%%%%%%%%%%%%%%%%%%%%%%%%%%
\subsection{Deformed conifold}
\lab{sec:Deformed}
%%%%%%%%%%%%%%%%%%%%%%%%%%%%%%%%%%%%%%%%%%%%%%%%%%%%%

There also exists the deformed conifold solution, which uses another method to rectify the singularity of the singular conifold \cite{Candelas:1989js,Klebanov:2000hb,Gwyn:2007qf,Ohta:1999we}. It involves expanding the singularity to the form of an $S^3$. This is also described by a parameter, one which defines the radius of the $S^3$, called $\epsilon$.

Again using the conventional left-invariant one-forms which satisfy (\ref{eq:2forms}), and defining $K$ by
\be
K=\frac{\left(\sinh(2r)-2r\right)^\frac{1}{3}}{2^\frac{1}{3}\sinh(r)},
\ee
we can write the deformed metric as:
\newcommand{\webvector}[4]{\left[ \begin{array}{c} #1 \\ #2 \\ #3 \\ #4 \end{array} \right]}

%\be
%\lab{eq:defMetric}
%ds^2\nn=-dt^2+\frac{\epsilon^{\frac{4}{3}}K}{2}\webvector{+\frac{1}{3K^3}dr^2}{+\frac{1}{2}\sinh^2\left(\frac{r}{2}\right)\left(\left(\sigma_1+\Sigma_1\right)^2+\left(\sigma_2-\Sigma_2\right)^2\right)}{+\frac{1}{2}\cosh^2\left(\frac{r}{2}\right)\left(\left(\sigma_1-\Sigma_1\right)^2+\left(\sigma_2+\Sigma_2\right)^2\right)}{+\frac{1}{3K^3}\left(\sigma_3+\Sigma_3\right)^2}
%\ee

\ba
\lab{eq:defMetric}
ds^2=-dt^2&+&\frac{\epsilon^{\frac{4}{3}}K}{2}
\left[\frac{1}{3K^3}dr^2+\frac{1}{3K^3}\left(\sigma_3+\Sigma_3\right)^2\right.\\\nonumber
&~&+\frac{1}{2}\sinh^2\left(\frac{r}{2}\right)\left(\left(\sigma_1+\Sigma_1\right)^2+\left(\sigma_2-\Sigma_2\right)^2\right)\\\nonumber
&~&+\left.\frac{1}{2}\cosh^2\left(\frac{r}{2}\right)\left(\left(\sigma_1-\Sigma_1\right)^2+\left(\sigma_2+\Sigma_2\right)^2\right)\right].
\ea
% 
% \be
% \lab{eq:defMetric}
% ds^2\nn=-dt^2+\frac{\epsilon^{\frac{4}{3}}K}{2}\left[\frac{1}{3K^3}dr^2+\frac{1}{2}\sinh^2\left(\frac{r}{2}\right)\left(\left(\sigma_1-\Sigma_1\right)^2+\left(\sigma_2+\Sigma_2\right)^2\right)+\frac{1}{2}\cosh^2\left(\frac{r}{2}\right)\left(\left(\sigma_1+\Sigma_1\right)^2+\left(\sigma_2-\Sigma_2\right)^2\right)+\frac{1}{3K^3}\left(\sigma_3+\Sigma_3\right)^2\right]
% \ee

At the origin, this metric is also smooth as it does not degenerate to a point but to an $S^3$ \cite{Candelas:1989js,Minasian:1999tt}, this 3-sphere has replaced the origin and so removed the conical singularity.

In the limit of high $r$ values we can see that this tends to the same form as the singular conifold by noting that for large $r$
\be
\nn K \rightarrow \left(\frac{2}{e^r}\right)^{\frac{1}{3}},
\ee
and so for large $r$,
\bea
ds^2|_{r\rightarrow\infty}\nn\rightarrow-dt^2&+&\frac{\epsilon^{\frac{4}{3}}}{6}\left(\frac{e^r}{2}\right)^{\frac{2}{3}}dr^2\\
&+&\frac{3}{4}\epsilon^{\frac{4}{3}} \left(2e^{2r}\right)^{\frac{1}{3}}
 \left(\frac{1}{6}\left(\sigma_1^2+\Sigma_1^2+\sigma_2^2+\Sigma_2^2\right)+\frac{1}{9}\left(\sigma_3+\Sigma_3\right)^2\right).
\eea

This can be written, in terms of a new radial coordinate $\rho$, as
\bea
\rho^2&=&\frac{3}{4}\epsilon^{\frac{4}{3}} \left(2e^{2r}\right)^{\frac{1}{3}},\\
ds^2|_{r=\infty}\nn&=&-dt^2+d\rho^2+\rho^2\left(\frac{1}{6}\left(\sigma_1^2+\Sigma_1^2+\sigma_2^2+\Sigma_2^2\right)+\frac{1}{9}\left(\sigma_3+\Sigma_3\right)^2\right).
\eea

This clearly shows that the asymptotic forms of both the singular conifold and the deformed conifold are the same, they only differ sufficiently close to the origin.
%%%%%%%%%%%%%%%%%%%%%%%%%%%%%%%%%%%%%%%%%%%%%%%%%%%%%
\subsection{Conifold transitions}
\lab{sec:transitions}
%%%%%%%%%%%%%%%%%%%%%%%%%%%%%%%%%%%%%%%%%%%%%%%%%%%%%

The deformed conifold and resolved conifold are each dependent on a continuous parameter, which describes the size to which the new tip has been extended. The limiting case of these parameters gives the singular conifold. The common limit to these two manifolds allows for the transition (through the singular conifold) from a resolved conifold to the deformed conifold, despite their differing topologies \cite{Candelas:1989js,Lukas:2004du}; this is called the conifold transition. This transition can also occur in the opposite direction and it is these processes that we shall be investigating.

We performed two classes of numerical simulations, one to simulate the collapse of a deformed conifold and one to simulate the collapse of the resolved conifold. Following from the results of these investigations we also construct a static black hole solution that we expect will be the final state of such collapses.
%%%%%%%%%%%%%%%%%%%%%%%%%%%%%%%%%%%%%%%%%%%%%%%%%%%%%
\subsection{Initial momentum}
\lab{sec:Momentum}
%%%%%%%%%%%%%%%%%%%%%%%%%%%%%%%%%%%%%%%%%%%%%%%%%%%%%
For both cases of evolving a resolved or deformed conifold we know what the initial metric is, this leaves us with a choice for what the initial time derivative of the metric could be, its "momentum". Noting that the spatial part of the metrics have co-dimension one - they depend only on $r$ - we make the simplifying choice that the initial momentum also depends only on $r$. As pointed out in \cite{Bizon:2005cp} the fact that we are in higher dimensions means that such radial dependence still allows for gravitational waves, so the gravitational dynamics is not trivial. With this restriction of a radially dependent momentum the metric will remain co-dimension one. On top of this restriction we must also satisfy the constraint equations, known as the Hamiltonian and momentum constraints, given in the appendices.
Our intention was to evolve the regular conifold geometries towards the singular conifold of section \ref{sec:sing} which we did by choosing a particular profile for the initial momentum, parametrized by an amplitude for the momentum, $P$, and a lengthscale over which the momentum was excited, $r_0$; the precise functional form is given in the relevant sections. These two parameters were the factors we varied in order to examine the collapse, in particluar we find the area of the black hole horizon as a function of these parameters.
%%%%%%%%%%%%%%%%%%%%%%%%%%%%%%%%%%%%%%%%%%%%%%%%%%%%%
\section{The collapse of the resolved conifold}
\lab{sec:ResCol}
%%%%%%%%%%%%%%%%%%%%%%%%%%%%%%%%%%%%%%%%%%%%%%%%%%%%%
To perform the simulations we need a suitable form for the metric which is general enough to give a consistent time evolution of the equations of motion, but with enough symmetry that numerical simulation is possible. For both the deformed and resolved conifold simulations we evolve a metric with four functions, each of which has an associated momentum. The equations of motion for these systems are given in the appendix and are written in such a way as to resemble the ADM formalism\cite{Arnowitt:1962hi,Kelly:2001kj}.
%%%%%%%%%%%%%%%%%%%%%%%%%%%%%%%%%%%%%%%%%%%%%%%%%%%%%
\subsection{Metric}
\lab{sec:Resmetric}
%%%%%%%%%%%%%%%%%%%%%%%%%%%%%%%%%%%%%%%%%%%%%%%%%%%%%

If we were able to solve the system analytically, a suitable form for the metric could be
\bea
\lab{eq:ResGeneralMetric}
ds^2\nn=-dt^2+a^2(t,r)dr^2&+&b^2(t,r)(\sigma_1^2+\sigma_2^2)
                       \\&+&c^2(t,r)(\Sigma_1^2+\Sigma_2^2)\\
                      \nn&+&d^2(t,r)(\sigma_3+\Sigma_3)^2.
\eea

However these may not be the optimum functions to evolve numerically. In particular we note that if we take this form for the metric, then the initial conditions corresponding to the resolved conifold would mean that these profile functions contained singular regions, albeit co-ordinate singularities.
In order to improve numerical stability, and to simplify the application of boundary conditions, we instead follow the method of \cite{Butcher:2007zk} and extract out various factors such that the functions we evolve are initially finite and asymptote to constant values, moreover they are symmetric under $r\rightarrow -r$.
\bea
\lab{eq:ResEvolveMetric}
ds^2\nn=-dt^2+A^2(t,r)dr^2&+&r^2B^2(t,r)(\sigma_1^2+\sigma_2^2)\\
                         &+&(6\alpha_0\,^2+r^2)C^2(t,r)(\Sigma_1^2+\Sigma_2^2)\\
                      \nn&+&r^2D^2(t,r)(\sigma_3+\Sigma_3)^2.
\eea

At $r=0$ we need to impose some boundary conditions, these follow by 
requiring local flatness at the origin \cite{Alcubierre:2004gn} and by maintaining $A$, $B$, $C$ and $D$ as even functions at the origin.

\bea
\lab{eq:resFlat}
A^0(t)&=&B^0(t)=C^0(t),\\
\nn\\
A(t,r)  &\sim& A^0(t)  +\mathcal{O} (r^2),\nn\\\lab{eq:resEven}
B(t,r)  &\sim& B^0(t)  +\mathcal{O} (r^2),\\\nn
C(t,r)  &\sim& C^0(t)  +\mathcal{O} (r^2),\\\nn
D(t,r)  &\sim& D^0(t)  +\mathcal{O} (r^2).\nn
\eea

%%%%%%%%%%%%%%%%%%%%%%%%%%%%%%%%%%%%%%%%%%%%%%%%%%%%%
\subsection{Initial conditions}
\lab{sec:Resinit}
%%%%%%%%%%%%%%%%%%%%%%%%%%%%%%%%%%%%%%%%%%%%%%%%%%%%%
By comparing (\ref{eq:ResEvolveMetric}) and (\ref{eq:resMetric}), we can read off the initial conditions for the resolved conifold,

\bea
\nn
A(0,r)  &=& \sqrt{\frac{r^2+6\alpha_0\,^2}{r^2+9\alpha_0\,^2}}\\
\label{eq:resConProf}
B(0,r)  &=& \frac{1}{\sqrt{6}},\\\nn
C(0,r)  &=& \frac{1}{\sqrt{6}},\\\nn
D(0,r)  &=& \frac{1}{3} \sqrt{\frac{r^2+9\alpha_0\,^2}{r^2+6\alpha_0\,^2}}.
\eea

This is a static metric, so if no initial momentum is added then no evolution occurs (this was used to test stability of our code). We added momentum and in doing so initiated the dynamical process, making sure that the momentum and Hamiltonian constraints were satisfied. Our specific algorithm was to add momentum to C (which describes the size of the two sphere) according to the form given below (\ref{eq:Cdot}). This depends upon two parameters giving the strength of the momentum ($P$) and the range the momentum extended away from the origin ($r_0$). 
The momentum for $B$ was taken as (\ref{eq:BDdot}), in order to simplfy the initial conditions, this actually has the effect of maintaining a 3-sphere at the origin.
\bea
\label{eq:Cdot}
\dot{C}(0,r)=-P\left(\frac{r}{r_0}\right)^4 e^{-\left(\frac{r}{r_0}\right)^2}\\
\label{eq:BDdot}
\frac{\dot{B}(0,r)}{B(0,r)}=\frac{\dot{D}(0,r)}{D(0,r)}.\label{eq:squashCond}
\eea

The form of the momentum is required to fall off sufficiently fast asymptotically so that we are not adding an infinite amount of energy, the exponential decay in (\ref{eq:Cdot}) achieves that and means we could have a final state of a finite mass black hole. The other two initial conditions for the $C$ and $D$ momenta were determined by the momentum constraint and the Hamiltonian constraint, see the appendix for the equations. These initial conditions, along with the Einstein vacuum equations are enough to evolve the system forward in time. This was achieved in practice with a fourth order Runge-Kutta scheme, where the time step was changed dynamically such that:
\be
\Delta t=0.005\lim_{r\to\infty}A(t,r)\; \Delta r 
\ee

We also used a fixed spatial step of 0.004 with a total of 6400 grid points. Throughout the simulation we monitored both the Hamiltonian constraint and the momentum constraint, finding that they remained below 0.01.

%%%%%%%%%%%%%%%%%%%%%%%%%%%%%%%%%%%%%%%%%%%%%%%%%%%%%
\subsection{Apparent horizons}
\lab{sec:ApHor}
%%%%%%%%%%%%%%%%%%%%%%%%%%%%%%%%%%%%%%%%%%%%%%%%%%%%%
We monitored the evolving system for the formation of an apparent horizon, a null trapped surface which implies the existence of an event horizon at some larger radius\cite{Thornburg:1995cp}.
The apparent horizon, unlike the event horizon, can be detected locally and does not require knowledge of the full future evolution of the metric, it is defined as a congruence of null geodesics that no longer increase their area. For radial null geodesics we may write,
\bea
0&=&\left[\frac{d\,Area}{dt}\right]_{null},\\
&=&\left[\frac{\partial \,Area}{\partial t}\right]_{r}
  +\left[\frac{\partial \,Area}{\partial r}\right]_{t}\,\left[\frac{{\rm d}r}{{\rm d}t}\right]_{null},
\eea
at the apparent horizon.
This can be tested for at each individual timeslice and so reveal the existence of an apparent horizon and, if one is found, imply the presence of a horizon.
We also measure the area of the apparent horizon at each time slice and so discover its evolution. 
For our simulations we find that its area increased monotonically but, as shown in Fig.  \ref{fig:ReshorArea},  asymptotes to a constant value. 
This constant value we took to be a good approximation to the area of the resulting event horizon.
%%%%%%%%%%%%%%%%%%%%%%%%%%%%%%%%%%%%%%%%%%%%%%%%%%%%%
\FIGURE{
\includegraphics[width=10cm]{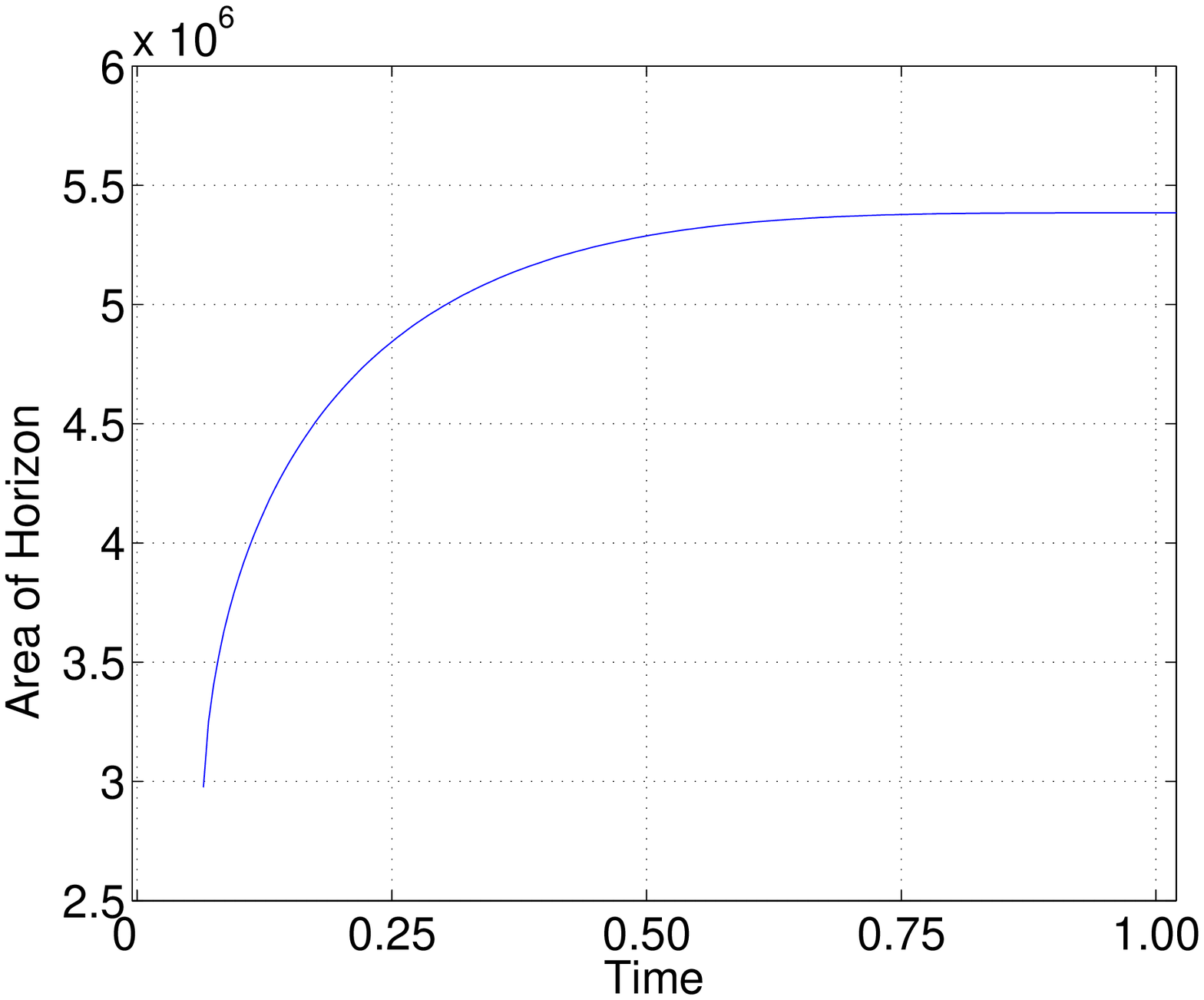}
\caption{The area of the apparent horizon of the collapsing resolved conifold, changing in time.} 
\lab{fig:ReshorArea}
}
%%%%%%%%%%%%%%%%%%%%%%%%%%%%%%%%%%%%%%%%%%%%%%%%%%%%%

%%%%%%%%%%%%%%%%%%%%%%%%%%%%%%%%%%%%%%%%%%%%%%%%%%%%%
\subsection{Results}
\lab{sec:Resresults}
%%%%%%%%%%%%%%%%%%%%%%%%%%%%%%%%%%%%%%%%%%%%%%%%%%%%%
We performed a number of simulations, varying the value of $P$ and $r_0$ in the initial conditions of
(\ref{eq:Cdot}), examined whether an apparent horizon formed, and measured its area if one did;
the results are presented in Fig. \ref{fig:ResRange}. For small values of $P$ and $r_0$ we were not able to reliably extract a value for the final area of the apparent horizon and so this portion of the plot has been left blank. This was due to the large timescale needed to produce such black holes, and was beyond the dynamic range of our simulations. What we see from the data is that for all the cases which could be reliably tested we did observe the formation of a black hole. This is in contradistinction to the results of the system examined in \cite{Butcher:2007zk} for the evolution of the collapsing cycle in the Eguchi-Hanson geometry.

%%%%%%%%%%%%%%%%%%%%%%%%%%%%%%%%%%%%%%%%%%%%%%%%%%%%%
\FIGURE{\includegraphics[width=10cm]{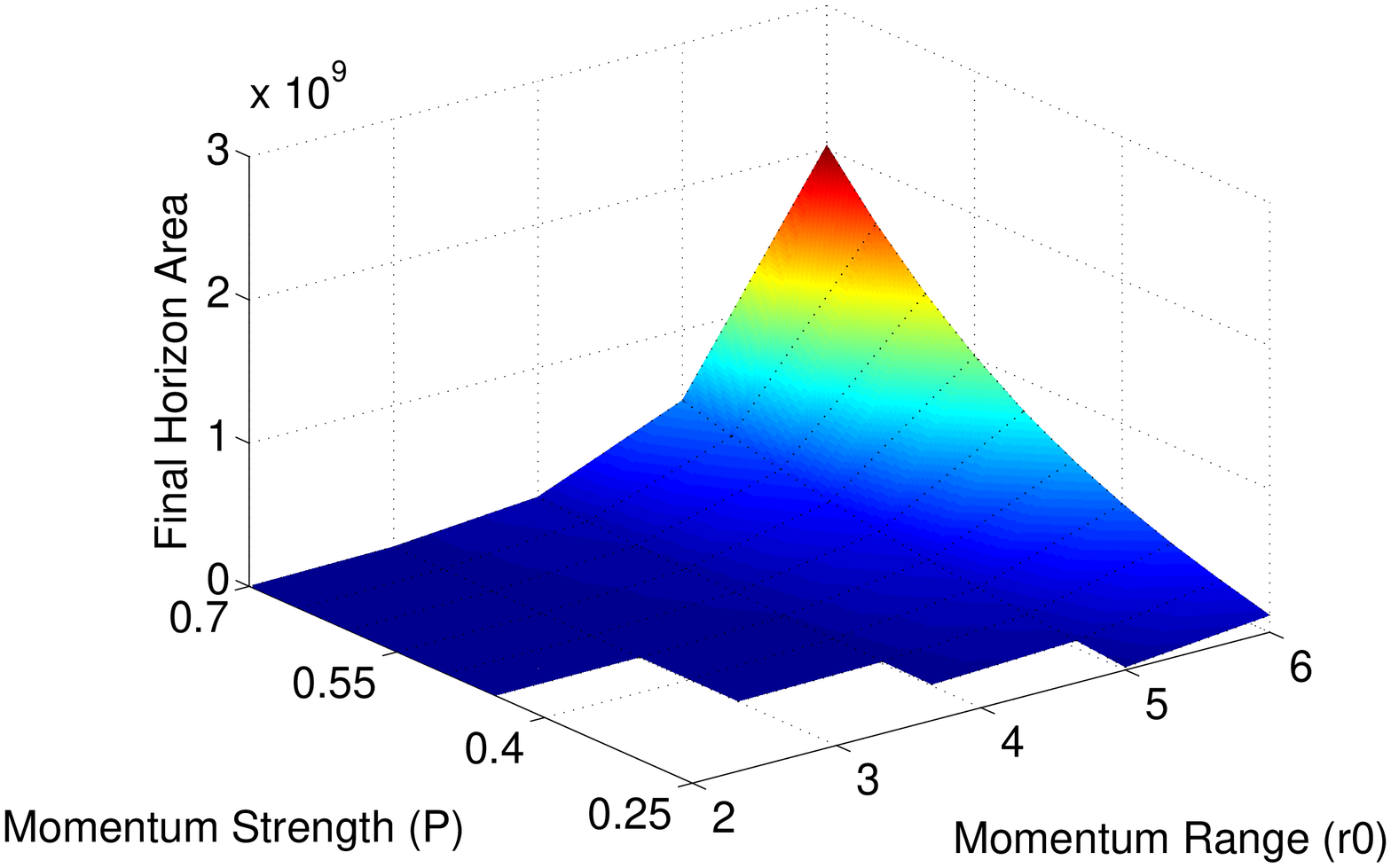}\includegraphics[width=8cm, height=5.5cm]{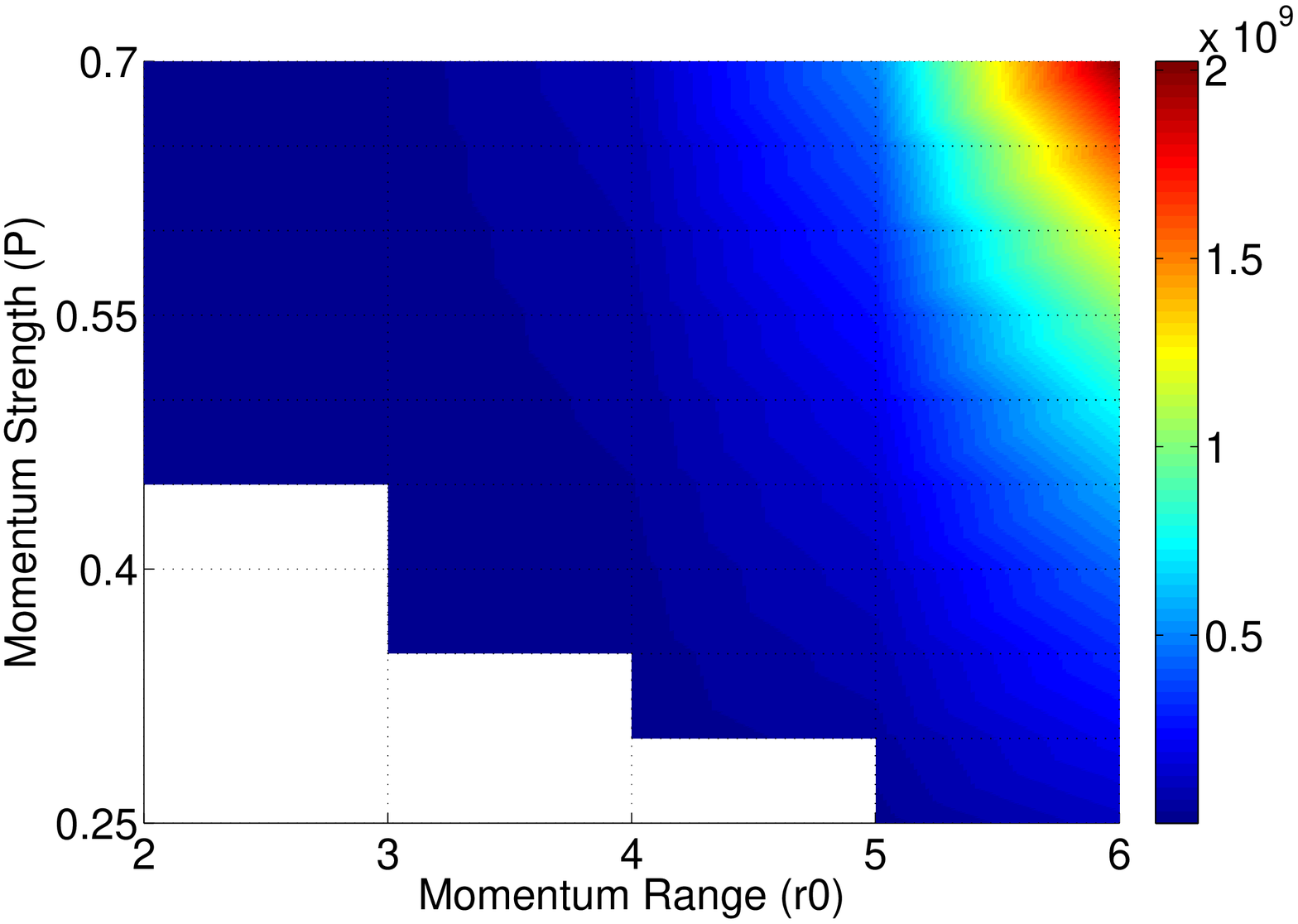}
\caption{Resolved Conifold: The dependence of the final area of the horizon on the initial momentum profile.} \lab{fig:ResRange}}
%%%%%%%%%%%%%%%%%%%%%%%%%%%%%%%%%%%%%%%%%%%%%%%%%%%%%

In addition to the area of the horizon another property of the resultant black hole is the squashing of the angular part of the metric. This exists because the parameters relating to $b$, $c$ and $d$ within (\ref{eq:ResGeneralMetric}) need not be equal at the horizon, and their various ratios are referred to as the squashings of the metric at the horizon. The parameters defining the extent of the squashing also change in time but, like the horizon area, converge as time goes on as shown in Fig.  \ref{fig:ResSquash} for a particular example. The fact that both the apparent horizon area, and the squashing at the horizon are asymtoting to constant values is evidence that the final state is settling on a static black hole, we shall present a candidate for these black holes in section \ref{sec:blackhole}.

%%%%%%%%%%%%%%%%%%%%%%%%%%%%%%%%%%%%%%%%%%%%%%%%%%%%%
\FIGURE{\includegraphics[width=10cm]{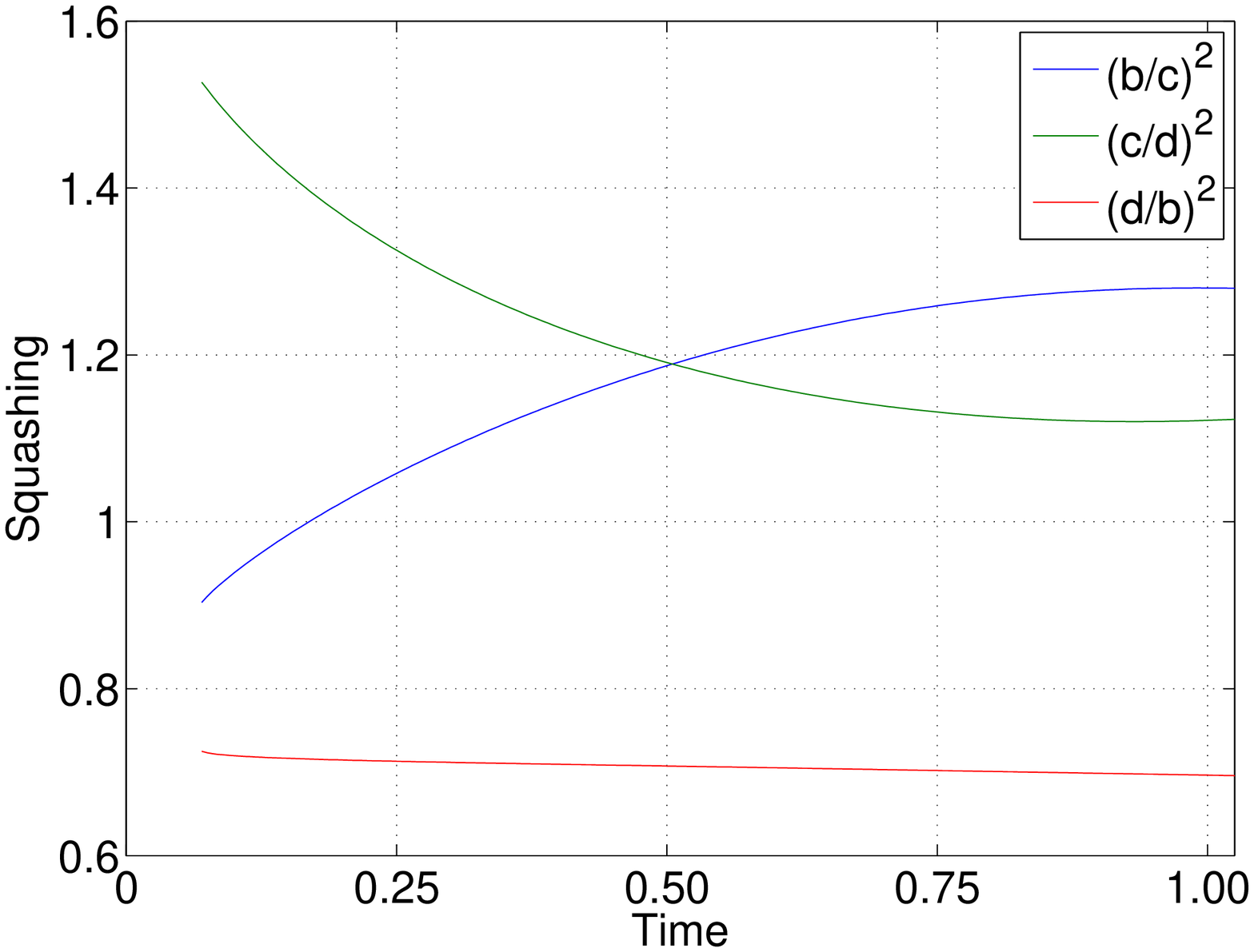}
\caption{The squashing values of (what was initially) the resolved conifold, seen to be converging in time. Note the slow change to the squashing (d/b) due to the initial condition (\ref{eq:squashCond})} \lab{fig:ResSquash}}
%%%%%%%%%%%%%%%%%%%%%%%%%%%%%%%%%%%%%%%%%%%%%%%%%%%%%

%%%%%%%%%%%%%%%%%%%%%%%%%%%%%%%%%%%%%%%%%%%%%%%%%%%%%
\subsubsection{Moduli space comparison}
\lab{sec:Modresults}
%%%%%%%%%%%%%%%%%%%%%%%%%%%%%%%%%%%%%%%%%%%%%%%%%%%%%
In section \ref{sec:Moduli} we presented a prediction for the behaviour of $\alpha(t)$ based on the moduli space approximation, given that the static resolved conifold is a solution for any fixed $\alpha$; this led us to conclude that $\sqrt{\alpha(t)}$ should evolve linearly in time. We would like to check this result against the full numerical solutions that we have obtained, but this involves some ambiguity in defining what we mean by $\alpha$ once the metric has evolved away from the precise form of the resolved conifold. That is to say, once the profile functions have evolved away from the functional form given by (\ref{eq:resConProf}), how do we extract a value for $\alpha$? In practise we chose to define $\alpha(t)$ using the value of $C(t,r)$ at the origin, $C(t,0)$ to extract a value for $\sqrt{\alpha(t)}$ by comparing expressions (\ref{eq:ResEvolveMetric}) and (\ref{eq:resMetric}),
\be
\sqrt \alpha(t)=\sqrt{6} \alpha_0 C(t,0).
\ee
 As shown in Fig. \ref{fig:ResModuli} the time evolution of $\sqrt{\alpha}$ is indeed linear in the initial stages.
%%%%%%%%%%%%%%%%%%%%%%%%%%%%%%%%%%%%%%%%%%%%%%%%%%%%%
\FIGURE{\includegraphics[width=10cm]{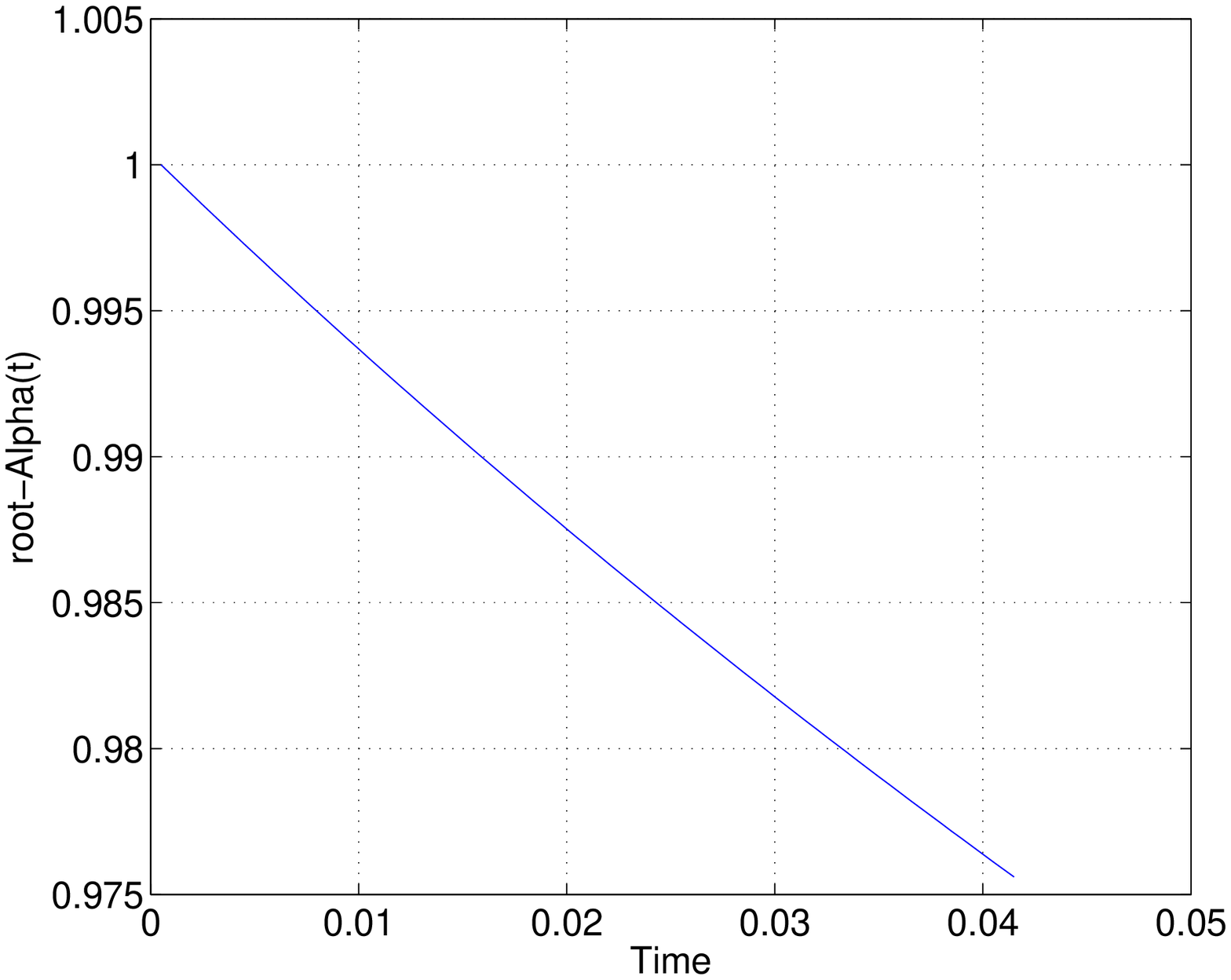}
\caption{The evolution of the the moduli of the resolved conifold in time. The approximation predicts this shall be linear.} \lab{fig:ResModuli}}
%%%%%%%%%%%%%%%%%%%%%%%%%%%%%%%%%%%%%%%%%%%%%%%%%%%%%

%%%%%%%%%%%%%%%%%%%%%%%%%%%%%%%%%%%%%%%%%%%%%%%%%%%%%
\section{The collapse of the deformed conifold}
\lab{sec:DefCol}
%%%%%%%%%%%%%%%%%%%%%%%%%%%%%%%%%%%%%%%%%%%%%%%%%%%%%
In the previous sections we described the collapse of the resolved conifold, that was understood to corresdpond to the contraction of the 2-sphere which was made finite in order to smooth out the conifold. Here we consider the second type of smoothing, namely the deformed conifold, which is achieved by having a finite size for the 3-sphere at the tip of the conifold. In the following sections we shall examine the time evolution of this geometry as this 3-sphere collapses.
%%%%%%%%%%%%%%%%%%%%%%%%%%%%%%%%%%%%%%%%%%%%%%%%%%%%%
\subsection{Metric}
\lab{sec:defmetric}
%%%%%%%%%%%%%%%%%%%%%%%%%%%%%%%%%%%%%%%%%%%%%%%%%%%%%
In (\ref{eq:defMetric}) we presented the metric for the deformed conifold, so a natural choice for the metric that we should evolve is
\bea
\lab{eq:defGeneralMetric}
ds^2\nn=-dt^2&+& a^2 dr^2\\
&+&b^2\left(\left(\sigma_1+\Sigma_1\right)^2+\left(\sigma_2-\Sigma_2\right)^2\right)\\\nn
&+&c^2\left(\left(\sigma_1-\Sigma_1\right)^2+\left(\sigma_2+\Sigma_2\right)^2\right)\\\nn
&+&d^2\left(\sigma_3+\Sigma_3\right)^2.
\eea

However, just as in the resolved case, this is not well-suited to a numerical solution owing to the profile functions diverging for the deformed conifold geometry. This is remedied by explicitly extracting the problematic terms, leaving us with smooth, finite functions to evolve, as well as allowing us to make the profiles symmetric under $r\rightarrow -r$. We therefore take the simulation metric to be
\bea
\lab{eq:defEvolveMetric}
ds^2\nn=-dt^2&+&\left(A(t,r) \cosh\left(\frac{r}{3}\right)\right)^2 dr^2\\
&+&\left(B(t,r) \sinh\left(\frac{r}{3}\right)\right)^2\left(\left(\sigma_1+\Sigma_1\right)^2+\left(\sigma_2-\Sigma_2\right)^2\right)\\\nn
&+&\left(C(t,r) \cosh\left(\frac{r}{3}\right)\right)^2\left(\left(\sigma_1-\Sigma_1\right)^2+\left(\sigma_2+\Sigma_2\right)^2\right)\\\nn
&+&\left(D(t,r) \cosh\left(\frac{r}{3}\right)\right)^2\left(\sigma_3+\Sigma_3\right)^2
\eea
so that, $A$, $B$, $C$ and $D$ are all even at the origin and also they asymptote to constant values. This allows for more accurate application of boundary conditions arising due to maintaining even variables at the origin (implying (\ref{eq:defEven})) and from requiring local flatness at the origin as described in \cite{Alcubierre:2004gn} (leading to (\ref{eq:defFlat})).

\bea
\lab{eq:defEven}
A(t,r)  &\sim& A^0(t)  +\mathcal{O} (r^2),\nn\\
B(t,r)  &\sim& B^0(t)  +\mathcal{O} (r^2),\\
C(t,r)  &\sim& C^0(t)  +\mathcal{O} (r^2),\nn\\
D(t,r)  &\sim& D^0(t)  +\mathcal{O} (r^2),\nn\\\nn
\\
\lab{eq:defFlat}
A^0(t)&=&B^0(t),\,\,\,\,\,C^0(t)=D^0(t)
\eea

%%%%%%%%%%%%%%%%%%%%%%%%%%%%%%%%%%%%%%%%%%%%%%%%%%%%%
\subsection{Initial conditions}
\lab{sec:definit}
%%%%%%%%%%%%%%%%%%%%%%%%%%%%%%%%%%%%%%%%%%%%%%%%%%%%%
By comparing (\ref{eq:defEvolveMetric}) and (\ref{eq:defMetric}), we can read off the initial conditions to be
\bea
\nn
A(0,r)^2  &=& \frac{\epsilon^{\frac{4}{3}}}{6 K^2 \cosh^2\left(\frac{r}{3}\right)}\\
B(0,r)^2  &=& \frac{1}{4} \epsilon^{\frac{4}{3}} K\frac{\sinh^2\left(\frac{r}{2}\right)}{\sinh^2\left(\frac{r}{3}\right)},\\\nn
C(0,r)^2  &=& \frac{1}{4} \epsilon^{\frac{4}{3}} K\frac{\cosh^2\left(\frac{r}{2}\right)}{\cosh^2\left(\frac{r}{3}\right)},\\\nn
D(0,r)^2  &=&  \frac{ \epsilon^{\frac{4}{3}}}{6 K^2 \cosh^2\left(\frac{r}{3}\right)}.
\eea

As these are derived from the static deformed conifold these initial conditions will yield a static metric, as was indeed found in our simulations by way of a check for the code. We added momentum of the following form
\be
\label{eq:momDef}
\frac{\dot{C}(0,r)}{C(0,r)}=\frac{\dot{D}(0,r)}{D(0,r)}=-P\left(\frac{r}{r_0}\right)^4 e^{-\left(\frac{r}{r_0}\right)^2},
\ee
to initiate the dynamics, with the momenta for the $A$ and $B$ functions being determined by the constraint equations in the appendix. By choosing the $C$ and $D$ momenta to related we are making the choice of maintaining the form of the 3-sphere, at least initially. As in the resolved conifold initial conditions we note that the momentum decays away sufficiently fast that the energy input is finite, allowing for the possibility of forming a finite mass black hole.

%%%%%%%%%%%%%%%%%%%%%%%%%%%%%%%%%%%%%%%%%%%%%%%%%%%%%
\subsection{Results}
\lab{sec:defResults}
%%%%%%%%%%%%%%%%%%%%%%%%%%%%%%%%%%%%%%%%%%%%%%%%%%%%%
We used the same methods as described in section \ref{sec:ApHor} to search for the appearance of an apparent horizon, thus indicating the presence of a horizon. The results are qualitatively similar to the collapse of the resolved conifold, with an apparent horizon appearing in all the cases we examined. These apparent horizons had a time development similar to Fig. \ref{fig:ReshorArea} and so we were able to establish a final value for their area in most cases, which we take to be a good approximation to area of the black hole horizon, this was checked by plotting the paths of outgoing null rays. The final area of the black hole depends upon the initial momentum we have added, which we characterized by a magnitude $P$ and a spatial extent $r_0$ in (\ref{eq:momDef}); varying these produced the graph displayed in Fig. \ref{fig:DefRange}. For small values of $P$ or $r_0$ the area of the apparent horizon did not converge sufficiently rapidly to aquire an accurate value for the horizon area, so we have left such regions blank in Fig. \ref{fig:DefRange} however, for all cases an apparent horizon was observed to form.

%%%%%%%%%%%%%%%%%%%%%%%%%%%%%%%%%%%%%%%%%%%%%%%%%%%%%
\FIGURE{\includegraphics[width=8cm, height=5.5cm]{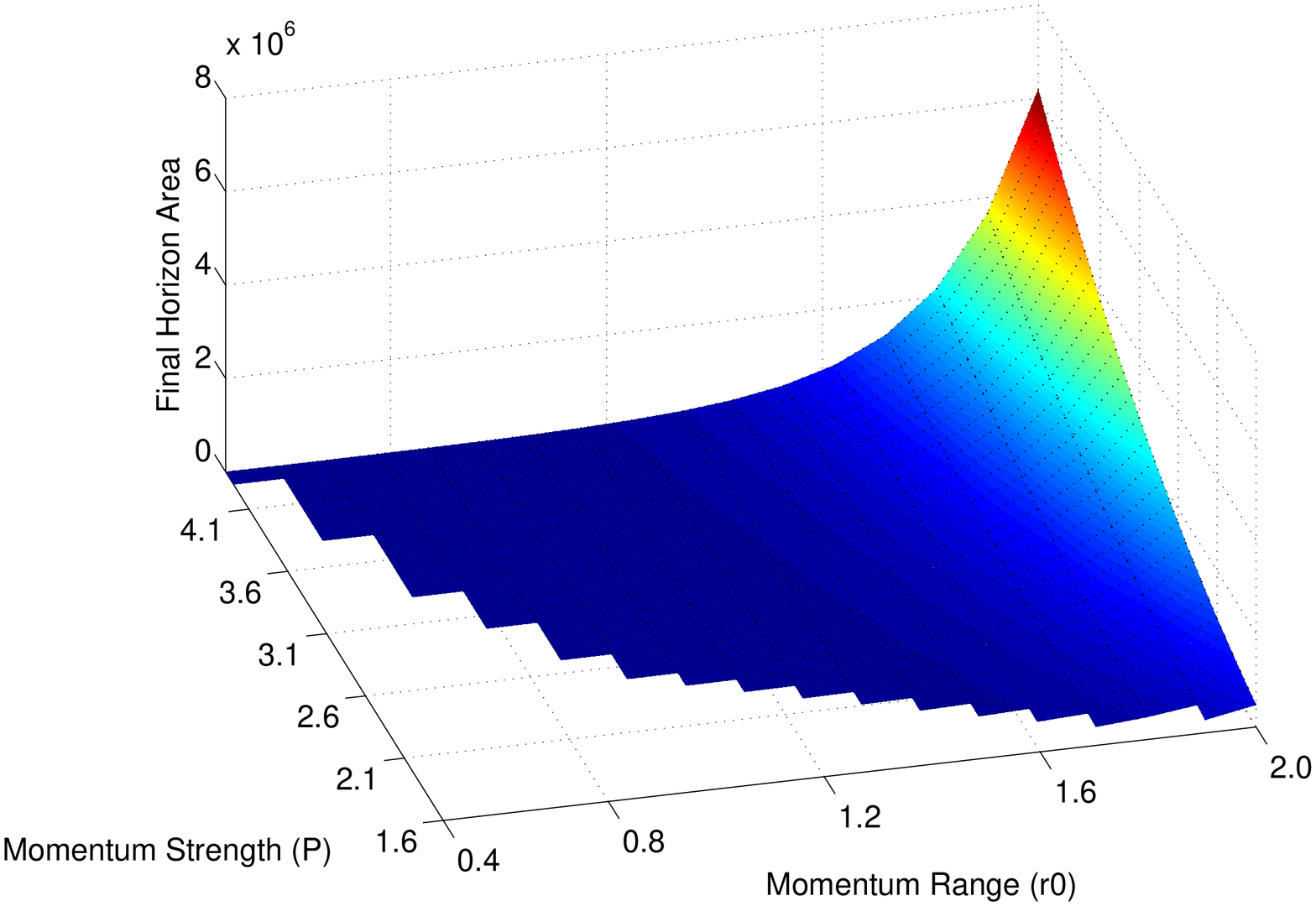}\includegraphics[width=8cm, height=5.5cm]{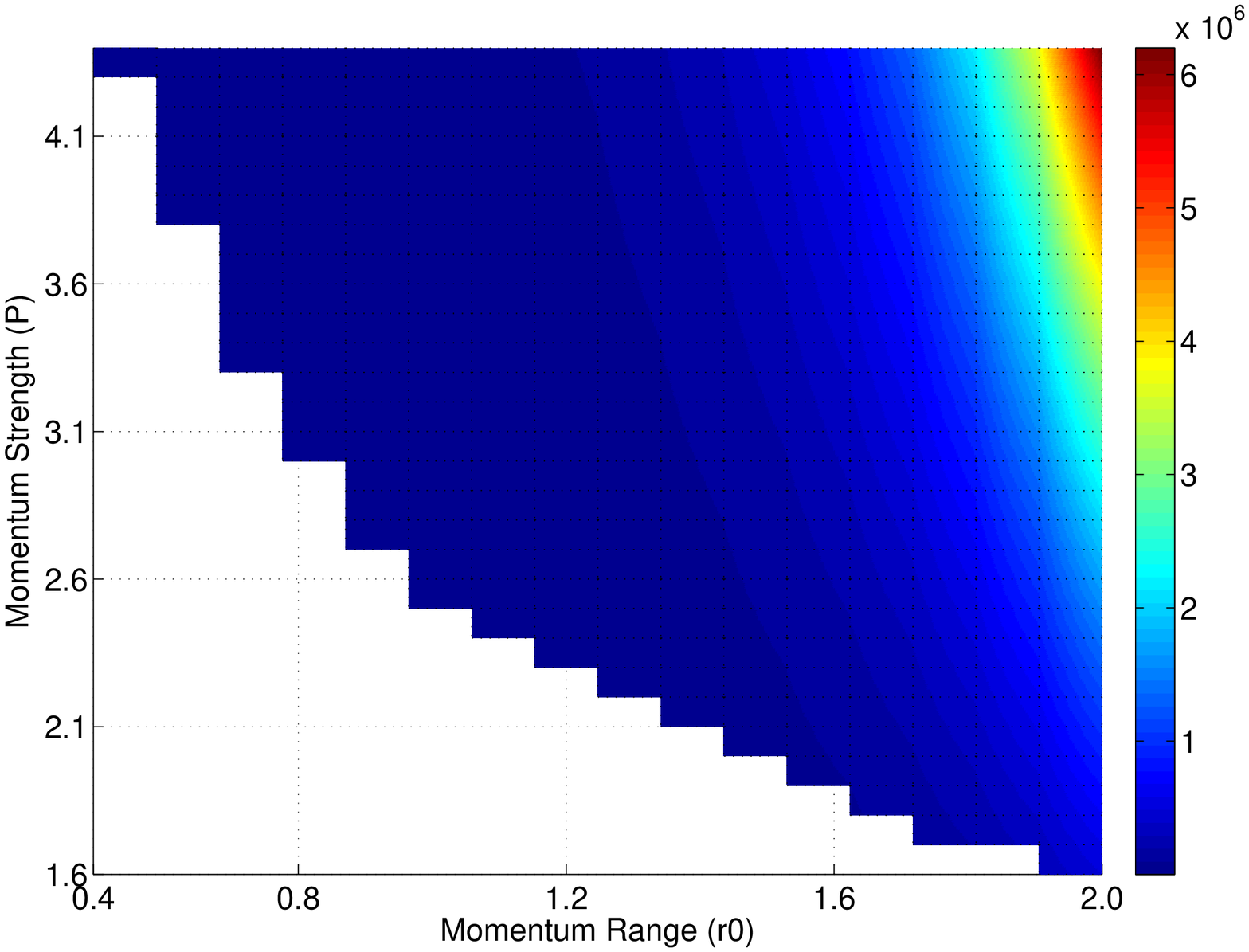}
\caption{Deformed Conifold: The effect on the final area of the horizon due to differing the initial momentum.} \lab{fig:DefRange}}
%%%%%%%%%%%%%%%%%%%%%%%%%%%%%%%%%%%%%%%%%%%%%%%%%%%%%

%%%%%%%%%%%%%%%%%%%%%%%%%%%%%%%%%%%%%%%%%%%%%%%%%%%%%
\section{The resulting black hole}
\lab{sec:blackhole}
%%%%%%%%%%%%%%%%%%%%%%%%%%%%%%%%%%%%%%%%%%%%%%%%%%%%%

For both the case of a collapsing resolved conifold and a collapsing deformed conifold we have seen that an apparent horizon always forms, for our choice of initial momentum, and this therefore implies the presence of a horizon. We have also seen that quantities such as the area of the apparent horizon, and the squashing modes of the angular part settle down to a constant value by the end of the simulation. This suggests that the final state of such collapses is a static black hole; in this section we suggest a possible candidate for the resultant black hole.

We suggest a metric of the form
\bea
\lab{eq:statMetric}
ds^2\nn=-f(r)dt^2+\frac{1}{f(r)}dr^2&+&b(r)^2(\sigma_1^2+\sigma_2^2)\\&+&c(r)^2(\Sigma_1^2+\Sigma_2^2)\\\nn&+&d(r)^2(\sigma_3+\Sigma_3)^2,
\eea
as a candidate for the final state. So starting from this ansatz we found the equations for the profile functions by requiring Ricci flatness, these are given in the Appendix \ref{app:static}
Unfortunately we were unable to derive an analytic solution to these equations, but we have been able to solve them numerically, showing that sensible solutions exist, one such solution with an event horizon at $r=1$ is presented in  Fig. \ref{fig:staticBH}. At the location of the event horizon we have that $f$ vanishes, with $df/dr$ finite and non-zero, showing that it is simply a co-ordinate singularity. At the horizon we see that the profile functions $b$, $c$ and $d$ are all different, indicating that the horizon has a squashed angular geometry. We also see that the profile function, $d$, associated to the U(1) direction $\sigma_3+\Sigma_3$ tends to a constant, rather than increasing linearly as $b$ and $c$ do. This shows that the spatial section of the black hole is not asymptotically a conifold. This behaviour was also seen in the case of collapsing cycles in the Eguchi-Hanson geometry\cite{Butcher:2007zk}, where a U(1) direction is picked out and aquires a constant radius asymptotically.
%%%%%%%%%%%%%%%%%%%%%%%%%%%%%%%%%%%%%%%%%%%%%%%%%%%%%
\FIGURE{\includegraphics[width=10cm]{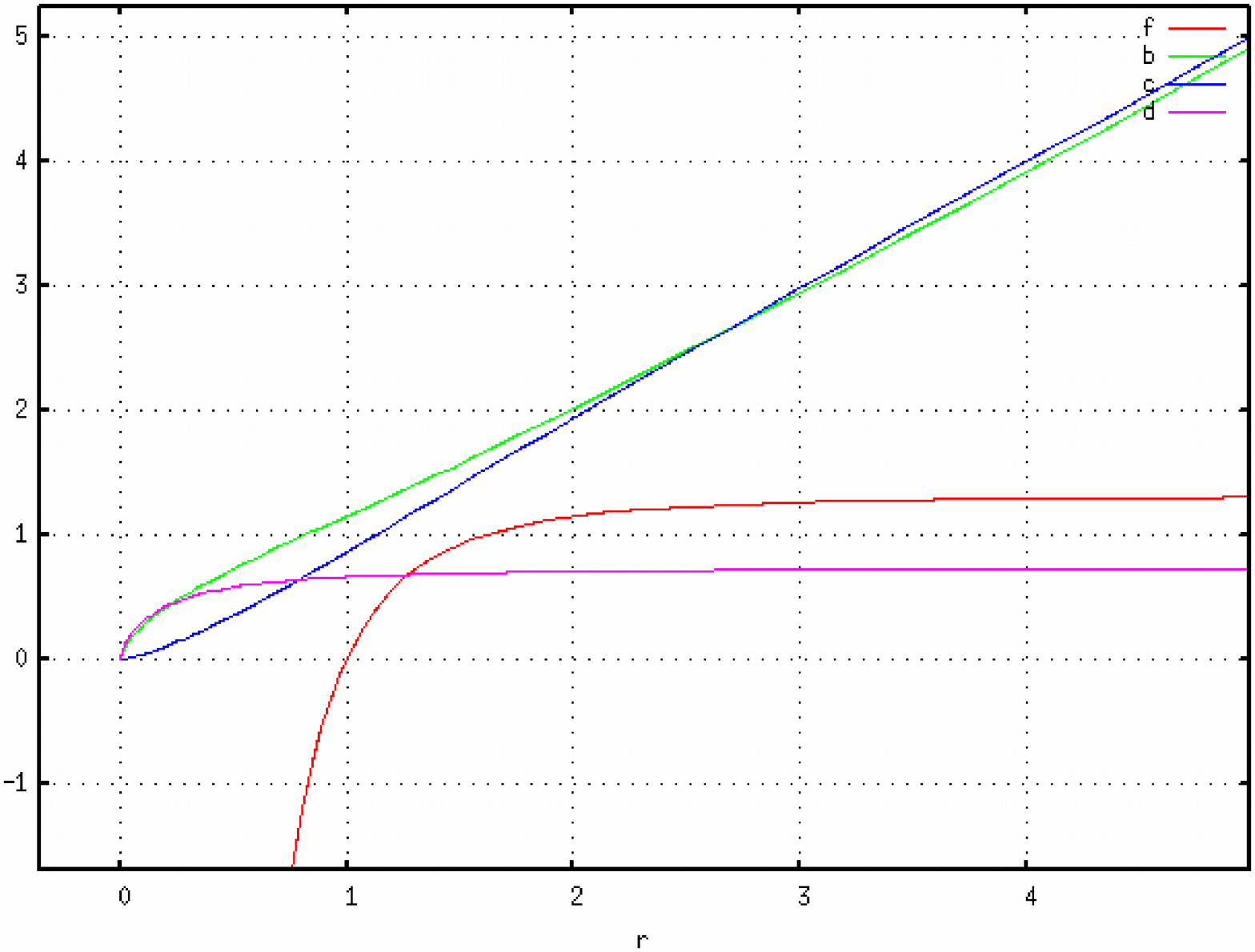}
\caption{The defining functions of the static black hole, with horizon conditions $r=1,f'=4,f=0,b=\half\sqrt{3}$ and $d=\frac{2}{3}$.} \lab{fig:staticBH}}
%%%%%%%%%%%%%%%%%%%%%%%%%%%%%%%%%%%%%%%%%%%%%%%%%%%%%

These are candidates for the black holes formed by the collapse of the resolved conifold or, with appropriate coordinate transformations, the deformed conifold. We should of course be careful about the asymptotic structure of the solution which should not be altered by the collapse process near the origin. The discrepancy between the initial condition being asymptotically conifold and the black hole asymptotics is resolved because the collapse takes an infinite amount of time to become the black hole, with a wave propogating outward from the collapse such that the interior is given by the black hole, and the exterior by the conifold.

%%%%%%%%%%%%%%%%%%%%%%%%%%%%%%%%%%%%%%%%%%%%%%%%%%%%%
\section{Conclusions}
\lab{sec:Conc}
%%%%%%%%%%%%%%%%%%%%%%%%%%%%%%%%%%%%%%%%%%%%%%%%%%%%%
Transitions capable of changing the topology of a Calabi-Yau manifold have been suggested previously \cite{Candelas:1988di,Candelas:1989ug,Greene:1996cy} and are a potential method for changing the topology of string vacua dynamically. These transitions would require the erstwhile distinct moduli spaces to share common, singular, manifolds. Dynamical application of the transitions would also require the cycles of the manifold to collapse successfully, that is to say without the formation of curvature singularities. Gravitational effects need to be accounted for as this collapse occurs, and failing to apply the effects of gravity may give false impressions for the transition process.
In order to study the processes leading to the collapse of a two sphere or a three sphere, we have performed numerical simulations of two different, but related, cases of collapse. The resolved conifold, which involves a collapsing two sphere, and the deformed conifold, which involves a collapsing three sphere. These spaces are formed by two different methods of smoothing out a singular conifold, and it is using this singular conifold geometry that allows string theory to join the moduli spaces of these two regular, distinct geometries\cite{Strominger:1995cz}.

In this first study of the gravitational dynamics of conifolds, beyond the moduli space approximation, we have found that horizons are formed by the collapse of cycles in either of the two regular conifold solutions. These horizons take the form of two sphere crossed with a squashed three sphere. Such a horizon formed in all the cases we studied, which is different to the situation found in \cite{Butcher:2007zk} where a study of the collapsing cycle in an Eguchi-Hanson geometry revealed cases where the cycle stopped collpasing, with no horizon being generated. However, the case of the conifold allows for a much larger choice of initial conditions, and the ubiquitous formation of a  horizon may be an artefact of our choice. 

Having established that horizons can form we have also found a possible candidate for the resulting black hole, by numerical means. This is a static black hole which also has a horizon comprising of a two sphere crossed with a squashed three sphere. An intriguing feature of this black hole, and one that it shares with the black associated to the collapse of the Eguchi-Hanson geometry, is that the asymtotic geometry contains a constant radius circle. 

In this paper we have only been interested in the gravitational dynamics coming from the Einstein-Hilbert action, and have not examined any stringy effects that may resolve the singularity. It is certainly true that if the scale at which the horizon forms is at the string scale, then these effects will be important. However, we have seen that a horizon forms for all of our initial conditions, so we should expect cases where the horizon size is greater than the string scale, and so field theory should be a valid approximation. For this initial study we have also neglected the dynamics from other supergravity fields such as the dilaton and form-fields. Indeed, this presents us with the natural extension of this work, namely to study the gravitational dynamics of the Klebanov-Strassler solution \cite{Klebanov:2000hb}. This solution represents a supersymmetric deformed conifold that is held in place by the presence of fluxes, and has found many applications in the realm of flux compactification \cite{Douglas:2006es} and brane inflation \cite{Kachru:2002kx}. Such calculations rely on the static Klebanov-Strassler geometry, without taking into account its own dynamics. If the dynamics of the Klebanov-Strassler geometry become singular then these calculations would need to be re-visited.

%%%%%%%%%%%%%%%%%%%%%%%%%%%%%%%%%%%%%%%%%%%%%%%%%%%%%
\acknowledgments 

NB would like to thank STFC for financial support.
%%%%%%%%%%%%%%%%%%%%%%%%%%%%%%%%%%%%%%%%%%%%%%%%%%%%%
% APPENDICES
\vskip 1cm
\appendix{\noindent\Large \bf Appendices}
%\renewcommand{\theequation}{\Alph{section}.\arabic{equation}}
%\setcounter{equation}{0}
%%%%%%%%%%%%%%%%%%%%%%%%%%%%%%%%%%%%%%%%%%%%%%%%%%%%%
\section{Resolved metric equations}
\lab{app:resolved}
%%%%%%%%%%%%%%%%%%%%%%%%%%%%%%%%%%%%%%%%%%%%%%%%%%%%%
The components of the Ricci tensor for the resolved conifold case (\ref{eq:ResGeneralMetric}) were calculated in the natural flat basis, and act to define both the equations of motion and the constraints. 
By defining the new functions
\bea
K_b=\frac{\dot{b}}{b}&\:&\:D_b=\frac{b'}{ab},\nn\\
K_c=\frac{\dot{c}}{c}&\:&\:D_c=\frac{c'}{cb},\nn\\
K_d=\frac{\dot{d}}{d}&\:&\:D_d=\frac{d'}{db},\nn\\
K_a=\frac{\dot{a}}{a},&\:&\:
\eea
we find
\bea
R_{11}=R_{22}&=&\dot{K_b}+K_b\left(K_a+2K_b+2K_c+K_d\right)-\frac{D_b'}{a}-D_b\left(2D_b+2D_c+D_d\right)+\frac{1}{b^2}-\frac{d^2}{2b^4},\nn\\
R_{33}=R_{44}&=&\dot{K_c}+K_c\left(K_a+2K_b+2K_c+K_d\right)-\frac{D_c'}{a}-D_c\left(2D_b+2D_c+D_d\right)+\frac{1}{c^2}-\frac{d^2}{2c^4},\nn\\
R_{55}&=&\dot{K_d}+K_d\left(K_a+2K_b+2K_c+K_d\right)-\frac{D_d'}{a}-D_d\left(2D_b+2D_c+D_d\right)+\frac{d^2}{2b^4}+\frac{d^2}{2c^4},\nn\\
R_{rr}&=&\dot{K_a}+K_a\left(K_a+2K_b+2K_c+K_d\right)-2\frac{D_b'}{a}-2\frac{D_c'}{a}-\frac{D_d'}{a}-\left(2D_b^2+2D_c^2+D_d^2\right),\nn\\
R_{tr}&=&2\left(\dot{D_b}+K_bD_b\right)+2\left(\dot{D_c}+K_cD_c\right)+\left(\dot{D_d}+K_dD_d\right),\nn\\
R_{tt}&=&-\left(\dot{K_a}+K_a^2+2\dot{K_b}+2K_b^2+2\dot{K_c}+2K_c^2\dot{K_d}+K_d^2\right).
\eea

Where the first four were used as evolution equations for the momenta, and the last two provided the constrint equations used to give consistent initial conditions, and to monitor the accuracy of the code.

%%%%%%%%%%%%%%%%%%%%%%%%%%%%%%%%%%%%%%%%%%%%%%%%%%%%%
\section{Deformed metric equations}
\lab{app:deformed}
%%%%%%%%%%%%%%%%%%%%%%%%%%%%%%%%%%%%%%%%%%%%%%%%%%%%%
The components of the Ricci tensor for the deformed conifold case (\ref{eq:defGeneralMetric}) were calculated in the natural flat basis, and act to define both the equations of motion and the constraints. 
By defining the new functions
\bea
K_b=\frac{\dot{b}}{b}&\:&\:D_b=\frac{b'}{ab},\nn\\
K_c=\frac{\dot{c}}{c}&\:&\:D_c=\frac{c'}{cb},\nn\\
K_d=\frac{\dot{d}}{d}&\:&\:D_d=\frac{d'}{db},\nn\\
K_a=\frac{\dot{a}}{a},&\:&\:
\eea
we find
\bea\nn
R_{11}=R_{22}&=&\dot{K_b}+K_b\left(K_a+2K_b+2K_c+K_d\right)-\frac{D_b'}{a}-D_b\left(2D_b+2D_c+D_d\right)+\frac{1}{2b^2}+\frac{b^2}{8c^2d^2}-\frac{c^2}{8b^2d^2}-\frac{d^2}{8c^2b^2},\nn\\
R_{33}=R_{44}&=&\dot{K_c}+K_c\left(K_a+2K_b+2K_c+K_d\right)-\frac{D_c'}{a}-D_c\left(2D_b+2D_c+D_d\right)+\frac{1}{2c^2}+\frac{c^2}{8b^2d^2}-\frac{b^2}{8c^2d^2}-\frac{d^2}{8c^2b^2},\nn\\
R_{55}&=&\dot{K_d}+K_d\left(K_a+2K_b+2K_c+K_d\right)-\frac{D_d'}{a}-D_d\left(2D_b+2D_c+D_d\right)+\frac{1}{2d^2}+\frac{d^2}{4b^2c^2}-\frac{b^2}{4c^2d^2}-\frac{c^2}{4d^2b^2},\nn\\
R_{rr}&=&\dot{K_a}+K_a\left(K_a+2K_b+2K_c+K_d\right)-2\frac{D_b'}{a}-2\frac{D_c'}{a}-\frac{D_d'}{a}-\left(2D_b^2+2D_c^2+D_d^2\right),\nn\\
R_{tr}&=&-2\left(\frac{K_b'}{a}+D_b(K_b-K_a)\right)-2\left(\frac{K_c'}{a}+D_c(K_c-K_a)\right)-\left(\frac{K_d'}{a}+D_d(K_d-K_a)\right),\\
R_{tt}&=&-\left(\dot{K_a}+K_a^2+2\dot{K_b}+2K_b^2+2\dot{K_c}+2K_c^2\dot{K_d}+K_d^2\right)
\eea

Where the first four were used as evolution equations for the momenta, and the last two provided the constrint equations used to give consistent initial conditions, and to monitor the accuracy of the code.

%%%%%%%%%%%%%%%%%%%%%%%%%%%%%%%%%%%%%%%%%%%%%%%%%%%%%
\section{Static black hole equations}
\lab{app:static}
%%%%%%%%%%%%%%%%%%%%%%%%%%%%%%%%%%%%%%%%%%%%%%%%%%%%%
The need for Ricci flatness imposes restriction upon the profile function of the metric (\ref{eq:statMetric}). These are calculated to be,

\bea
\nn
f''&=&-2f'\left(\frac{b'}{b}+\frac{c'}{c}+\frac{d'}{2d}\right)\\\nn
b''&=&-\frac{f'b'}{f}+\frac{4}{bf}-\frac{d^2}{2fb^3}-\frac{b'b'}{b}-\frac{2b'c'}{c}-\frac{b'd'}{d}\\
c''&=&-\frac{f'c'}{f}+\frac{4}{cf}-\frac{d^2}{2fc^3}-\frac{c'c'}{c}-\frac{2b'c'}{b}-\frac{c'd'}{d}\\\nn
d''&=&-\frac{f'd'}{f}+2\left(\frac{d^3}{4fb^4}-\frac{d'b'}{b}+\frac{d^3}{4fc^4}-\frac{d'c'}{c}\right)\\\nn
0&=&f'\left(\frac{2b'}{b}+\frac{2c'}{c}+\frac{d'}{d}\right)+\frac{d^2}{2b^4}+\frac{d^2}{2c^4}-\frac{8}{b^2}-\frac{8}{c^2}\\\nn
& &+f\left(2\left(\frac{b'}{b}\right)^2+2\left(\frac{c'}{c}\right)^2+8\frac{b'c'}{bc}+4\frac{b'd'}{bd}+4\frac{d'c'}{dc}\right).
\eea

\end{document}